\documentclass[aps,prl,showpacs,twocolumn,preprintnumbers,amsmath,amssymb]{revtex4-1}

\usepackage{graphicx}
\usepackage{dcolumn}
\usepackage{bm}
\usepackage{color}
\usepackage{soul}

\def\st#1{\unskip\relax}

\begin{document}

\title{
Dispersive dam-break flow of a photon fluid}

\author{Gang Xu}
\email{gang.xu@ircica.univ-lille1.fr} 
\author{Matteo Conforti}
\author{Alexandre Kudlinski}
\author{Arnaud Mussot}
\affiliation{
Univ. Lille, CNRS, UMR 8523 - PhLAM - Physique des Lasers Atomes et Mol{\'e}cules, F-59000 Lille, France
}

\author{Stefano Trillo}
\email{stefano.trillo@unife.it} 
\affiliation{Department of Engineering, University of Ferrara, Via Saragat 1, 44122 Ferrara, Italy}


\date{\today} 
 
\begin{abstract} 
We investigate the temporal photonic analogue of the dam-break phenomenon for shallow water by exploiting a fiber optics setup. We clearly observe the decay of the step-like input (photonic dam) into a pair of oppositely propagating rarefaction wave and dispersive shock wave. Our results show evidence for a critical transition of the dispersive shock into a self-cavitating state. The detailed observation of the cavitating state dynamics allows for a fully quantitative test of the Whitham modulation theory applied to the universal defocusing nonlinear Schr\"odinger equation.
\end{abstract} 
 

 
\maketitle 

{\em Introduction.---} 
The laser light propagating in nonlinear media often behaves as a photon fluid \cite{Brambilla91,Onorato13PR,RTinst,Turitsyna13,Wan07,inchshock,Carusotto13,Vocke16}, sharing phenomena that characterise fluid flows such as rogue waves \cite{Onorato13PR}, instabilities \cite{RTinst}, transition to turbulence \cite{Turitsyna13}, coherent \cite{Wan07} and incoherent \cite{inchshock} shock waves, and superfluid flow around obstacles \cite{Carusotto13,Vocke16}. 
In regimes described by the defocusing nonlinear Schr\"odinger equation (NLSE), a distinctive trait of the photon fluid evolution is the formation of dispersive shock waves (DSWs, or undular bores) \cite{Wan07,Conti09,Fatome14,Wang15,Xu16,Millot16}, fast oscillating wavetrains that spontaneously emerge from the tendency to develop a gradient catastrophe \cite{ElHoefer16,Grava,Kamchatnov}. DSWs are ubiquitous, being observed in other systems ruled by the NLSE such as cold atom condensates \cite{Hoefer06} and spin waves \cite{spin}, as well as in other dispersive hydrodynamic settings involving, e.g. electrons \cite{Mo13}, water waves \cite{Trillo16} and viscous fluid conduits \cite{Maiden16}.
A major breakthrough in the analytical description of the DSW is the Whitham modulation theory \cite{ElHoefer16,Grava,Kamchatnov,Gurevich_JETP87,El_PhysD95}, which, however, assumes the DSW to develop from step-like initial conditions (the Riemann problem), whereas experiments to date have been mostly concerned with smooth or periodic initial conditions. 
Therefore experiments devoted to investigate the dispersive Riemann problem are of paramount importance for advancing the understanding of dispersive hydrodynamic flows, a fortiori for systems such as the NLSE where the modulation theory predict critical transitions in the behavior of the shock \cite{El_PhysD95,Hoefer06}.\\
\indent In this letter, we exploit a fiber optics set-up to investigate experimentally the Riemann problem associated with an initial step, in the temporal domain, in the optical power. In the absence of any frequency chirp across the jump such a problem is isomorphic to the classic 1D dam-break problem of hydrodynamics \cite{Stoker,Whitham,Leveque}.
Indeed, we demonstrate that the light evolves as a fluid mimicking the basic features of the dam-break in shallow water, namely the decay into a shock and a rarefaction-wave (RW) pair, connected by an expanding plateau. The dispersive character of the shock, however, leads to a critical transition, which is predicted in the framework  of Whitham theory \cite{El_PhysD95,Hoefer06}. Above a critical height of the jump, we report evidence for the onset of self-cavitation, i.e. the appearance of a null point in the optical power. The full experimental characterisation of the cavitating state allows, for the first time, for a quantitative comparison with modulation theory.

\begin{figure}[htb]
\centering
\includegraphics[width=8.3cm]{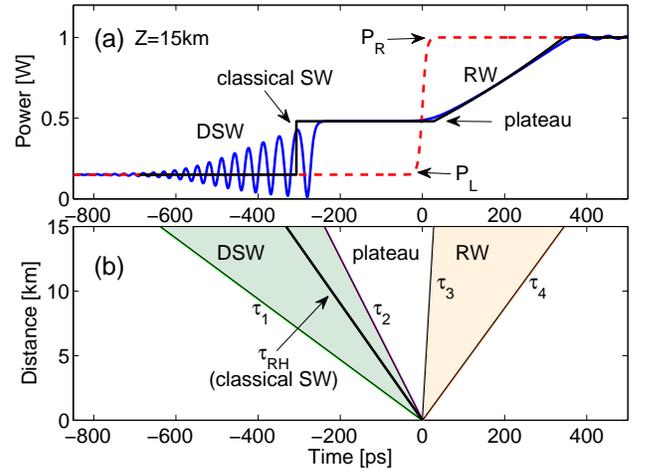}
     \caption{
     (a) Snapshots comparing the DSW-RW pair (solid blue curve) obtained from the NLSE with step-like input $E(T,0)=[P_L+(P_R-P_L)(1+\tanh(T/T_r))/2]^{1/2}$ (dashed red) with the ideal dispersionless solution of the SWEs (solid black). The oscillations over the top of the RW are due to the Gibbs phenomenon \cite{Gibbs}. 
     (b) Wedges in $(T,Z)$ plane corresponding to the RW (fable orange), the DSW (green) and the plateau in between. The boundaries correspond to slopes $\sqrt{k'' \gamma P_R} \tau_j$, $j=4,3,2,1$. Here $k''=176$ ps$^2$/km and $\gamma=3$ (W km)$^{-1}$, $P_R=1$ W, $P_L=150$ mW, $T_r=10$ ps.
}
\label{fig1}
\end{figure}

{\em Theory of dispersive dam-break.---} The dam-break Riemann problem that we investigate is described by the NLSE that rules the evolution of the temporal envelope field $E(T,Z)$ along the fiber distance $Z$
\begin{equation} \label{nls}
i  \frac{\partial E}{\partial Z} - \frac{k''}{2}\frac{\partial^{2} E}{\partial T^{2}} +  \gamma \left| E \right|^{2} E = 0,
\end{equation}
subject to a step initial condition in $T=0$, i.e. $E(T,0)=\sqrt{P_L}$ for $T<0$, and $E(T,0)=\sqrt{P_R}$ for $T>0$, where the constant left and right power levels $P_L$ and $P_R \ge P_L$ define the bottom and the top of the optical dam in $T=0$ (see Fig. \ref{fig1}(a)).
Here $T=T_{lab}-k' Z$ is the retarded time in a frame moving with group-velocity $1/k'=dk/d\omega \vert_{\omega_0}^{-1}$, whereas $k''=d^2k/d\omega^2 \vert_{\omega_0}=176$ ps$^2$/km and $\gamma=3$ (W km)$^{-1}$ are the dispersion and the nonlinear coefficient of our fiber, $\omega_0$ being the carrier frequency.  $\gamma k''>0$ corresponds to the defocusing regime of the NLSE. 
The Madelung transform $E(T,Z)=\sqrt{P_R} \sqrt{\rho(t,z)} \exp \left(-i \int_{-\infty}^{t} u(t',z) dt' \right)$ allows to formulate the NLSE in hydrodynamical form:
\begin{eqnarray} 
& &\rho_z + (\rho u)_t = 0\;;\label{swe1} \\ 
& &u_z + \left( \frac{u^2}{2} + \rho \right)_t= \frac{1}{4}\left[ \frac{\rho_{tt}}{\rho} - \frac{(\rho_t)^2}{2\rho^2} \right]_t, \label{swe2}
\end{eqnarray}
where we set $z=Z/Z_{0}$, $t=T/T_0$ with $Z_{0} \equiv (\gamma P_R)^{-1}$ and $T_{0} \equiv \sqrt{k''/{\gamma P_R}}$. By neglecting the RHS containing higher-order derivatives (quantum pressure term \cite{Wan07}), Eqs. (\ref{swe1}-\ref{swe2}) constitute the dispersionless vector conservation law known as the shallow water equations (SWEs) \cite{Leveque}. The role of local water depth and longitudinal velocity are played, here, by the normalized power $\rho=|E|^2/P_R$ and chirp $u$, whereas space and time have interchanged role. The evolution of an initial ($z=0$) step elevation from $\rho_L$ to $\rho_R>\rho_L$ \cite{rho}, with $u(t,0)$ identically vanishing, is the classic dam-break Riemann problem. The solution of the SWEs, first given by Stoker \cite{Stoker}, can be formulated in terms of the self-similar variable $\tau=t/z$ \cite{warnvelocity} and involves a classical shock wave and a RW. As shown by the black solid line in Fig. \ref{fig1}(a), the shock and the RW propagate in opposite directions (towards $t<0$ and $t>0$, respectively, or down- and up-stream directions in the hydrodynamic problem), being connected by an expanding plateau characterised by intermediate constant values $\rho_i= (\sqrt{\rho_L} + \sqrt{\rho_R})^2/4$, $u_i= \sqrt{\rho_L} - \sqrt{\rho_R}$. The step values $\rho_L$ and $\rho_R$ also fix the edge velocities of the smooth RW to the values:
\begin{equation} 
\tau_4=\sqrt{\rho_R};\;\;\tau_3=\frac{3\sqrt{\rho_L}-\sqrt{\rho_R}}{2} \label{RW}.
\end{equation}
Conversely, the jump from $\rho_i, u_i$ to $\rho_L, u_L=0$ constitutes a classical shock which moves with velocity $\tau_{RH}=\frac{\rho_i u_i}{\rho_i - \rho_L}$, deriving from the well-known Rankine-Hugoniot condition \cite{Leveque}.
However, such shock is regularized into an oscillating DSW by the effect of the RHS of Eq. (\ref{swe2}) that stems from dispersion.
A snapshot of the breaking scenario obtained from the full NLSE is compared with the dispersionless limit (SWEs) in Fig. \ref{fig1}(a). The DSW is delimited by two edge velocities $\tau_{1,2}$ ($ \tau_1< \tau_{RH} < \tau_2$), where the oscillations vanish (linear edge) or become deepest (soliton edge), respectively \cite{El_PhysD95,Hoefer06,ElHoefer16}.
According to Whitham modulation theory such velocities reads as  \cite{El_PhysD95,supplementary}
\begin{equation}
\tau_2=-\frac{\sqrt{\rho_L}+\sqrt{\rho_R}}{2}; \;\;\tau_1=\frac{\rho_L-2\rho_R}{\sqrt{\rho_R}},\label{DSW}
\end{equation}
whereas, in the same framework, owing to the RW smoothness, one recovers Eqs. (\ref{RW}) for the RW edges. The velocities $\tau_{1,2}$ in Eqs. (\ref{DSW}) and $\tau_{3,4}$  in Eqs. (\ref{RW}) define the wedges where the DSW and RW expand, as shown (in dimensional units, i.e. including the multiplicative factor $T_0/Z_0=\sqrt{k'' \gamma P_R}$ for our fiber) in Fig. \ref{fig1}(b). Importantly, the modulation theory entails a crossover between two different regimes separated by a critical condition. In the first regime, the DSW envelopes are monotone and the DSW power never vanishes, as shown in Fig. \ref{fig1}(a). However, below a critical value of the key parameter, namely the ratio between the quiescent (down- and up-stream) states, that we henceforth denote as $r=\rho_L/\rho_R=P_L/P_R$, the DSW exhibits a self-cavitating point (i.e., zero power, corresponding to a vacuum point in gas dynamics), which we found at  \cite{supplementary}
\FL \begin{equation} \label{vacuum}
\tau_0 = \sqrt{\rho_L} - \sqrt{\rho_R} + 2\sqrt{\rho_L} \left[1 -  \frac{\sqrt{\rho_R} - \sqrt{\rho_L}}{\sqrt{\rho_R} - 3\sqrt{\rho_L}} \frac{E(m)}{K(m)} \right]^{-1},
\end{equation}
where $E$ and $K$ are elliptic integrals of the first and second kind, respectively,  of modulus $m =4\rho_L/(\sqrt{\rho_L} - \sqrt{\rho_R})^2$.   
The threshold for cavitation can be obtained by imposing $m=1$ (cavitation on soliton edge of the DSW), or equivalently $\tau_2=u_i$, from which we obtain
\begin{equation} \label{th}
r_{th} \equiv \left( \frac{\rho_L}{\rho_R} \right)_{th} \equiv \left( \frac{P_L}{P_R} \right)_{th} = \frac{1}{9} \simeq 0.11.
\end{equation}
For $P_L/P_R$ below such threshold, a vacuum point always exists, which tends towards the linear edge of the DSW in the limit $P_L \rightarrow 0$. In this limit, however, both the oscillation amplitude and the plateau extension tends to vanish \cite{El_PhysD95}, and one recovers the limiting hydrodynamic case known as ``dry-bed" dam-break, characterised by a single RW extending to zero and no shock \cite{Whitham,drybedexp,Kodama95}.

\begin{figure}[t!]
\centerline{\includegraphics[width=8.3cm]{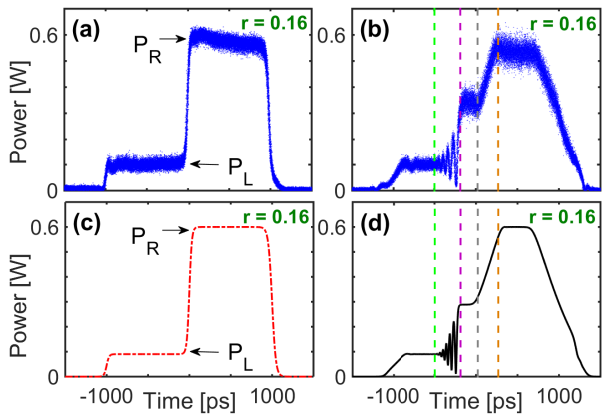}}
\caption{
Temporal traces of the whole waveform: (a,b) experiment; (c,d) numerics. The left column (a,c) shows the input; The right column (b,d) is the output power profile after propagation along the 15 km long fiber. The vertical dashed lines give the predicted delays of the edges of the RW (gray and orange lines, from Eqs. (\ref{RW})) and the DSW (magenta and green lines, from Eqs. (\ref{DSW})), respectively.
}
\label{fig2}
\end{figure}

{\em Experiment.---} 
We performed a series of experiments to provide evidence for the decay of a photonic dam into the DSW-RW pair and for the critical transition to self-cavitation.
In our experimental setup (see  \cite{supplementary}), we make use of a continuous laser diode source emitting at $\lambda=1560$ nm, which is intensity modulated by an electro-optic modulator driven by an arbitrary waveform generator, with typical rise time $T_r \sim 25$ ps (raised tanh shape, rising from $10 \%$ to $90 \%$ in $\sim 50$ ps) and 78 MHz repetition rate. The signal is pre-amplified in a semiconductor optical amplifier, and then it passes through a spectral filter to remove amplified spontaneous emission in excess, and is finally amplified by means of an Erbium-doped fiber amplifier. This signal is launched in a dispersion compensating fiber (parameters as in the caption of Fig. \ref{fig1})
and analysed in the time domain with an optical sampling oscilloscope with 1.6 ps resolution. A typical shape of the step-like signal that we obtained is shown in Fig. \ref{fig2}(a). Two elevating steps in power from zero to $P_L$ and from $P_L$ to $P_R$, can be adjusted independently, and are followed by a descending step (trailing edge) back to zero. The advantage of using this specific signal shape is that, in the same experimental run, we can compare (i) the DSW-RW dynamics developing around the jump from $P_L$ to $P_R$, with (ii) the ``dry-bed" dynamics developing over the descending step. Importantly, the duration of each of the constant power states $P_L,P_R$ was adjusted to be long, up to $\sim 1$ ns, so that the DSW-RW pair can develop without feeling the interaction with the first step and the trailing edge of the waveform. Clearly, as can be seen in Fig. \ref{fig2}(a), the main step ($P_L$ to $P_R$) is not instantaneous. However, the short rising time allows us to clearly observe the dam-break phenomenon, as we will see below. Another challenging issue faced in the experiment is the loss compensation. Indeed, the DSW-RW dynamics is very sensitive to losses and even weak losses of optical fibers (0.5 dB/km) are strongly detrimental. For instance, the plateau that connects the RW and the DSW would be completely distorted, not allowing for a quantitative comparison with theory (see Figs. S2 and S3 in \cite{supplementary}). Inspired from transparent telecommunication networks, we counterbalanced linear losses by means of Raman amplification \cite{AgrawalRaman}. To this end, a counterpropagating beam was injected in the fiber at $\lambda=1482$ nm. In this way, we achieve significant (close to peak) Raman gain with weak relative noise intensity transfer \cite{AgrawalRaman}.

Figure \ref{fig2}(b) shows the output temporal trace obtained for the input step-like signal with $P_R=0.6$ W, $P_L \simeq 100$ mW ($r \simeq 0.16$).  As can be seen, the input optical dam at $T=0$ breaks into a DSW-RW  pair. The DSW is characterized by fast oscillations with $\sim 40$ ps average period (the period scales proportionally to $\sqrt{k''/(\gamma P_i)}$). The RW smoothly connects the plateau with power $P_i \simeq 0.3$ W (in agreement with the theoretical prediction $P_i=P_R \rho_i = 0.297 $ W) to the peak level $P_R$. Conversely, over the trailing edge, no shock occurs and a smooth RW dropping to zero is observed, consistently with the case of ``dry-bed" dam-break. Overall, the data show an excellent agreement with the numerical simulations reported in Fig. \ref{fig2}(d) based on the NLSE (\ref{nls}). We emphasise that the occurrence of the DSW-RW pair is related to the non-zero background and not to the character (ascending vs. descending) of the step. Indeed the DSW-RW pair is observed on the trailing edge for a mirror-symmetric input.
 
\begin{center}
\begin{figure}[t!]
\includegraphics[width=8.5 cm]{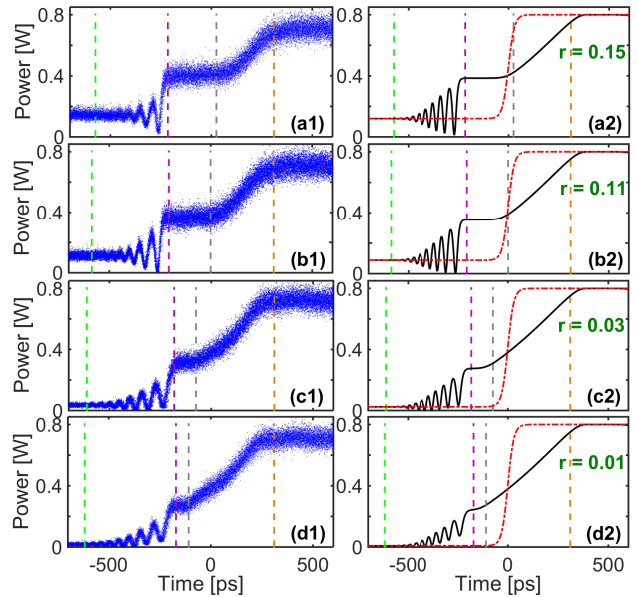}
\caption{Temporal traces at the fiber output for constant peak power of the photonic dam $P_R=0.8 W$ and different background fractions $r$: (a1,a2) $0.15$; (b1,b2) $0.11$; (c1,c2) $0.03$; (d1,d2) $0.01$. 
Left column: experimental results. Right column: numerical simulation based on the NLSE.}
\label{fig3}
\end{figure}
\end{center}

We have then proceeded to investigate in detail the breaking dynamics of the step-like input into the DSW-RW pair for different heights of the optical dam. The results shown in Fig. \ref{fig3}, are obtained for a fixed peak power $P_R=0.8$ W (slightly larger than that of Fig. \ref{fig2}), and a variable background $P_L$, i.e. a variable ratio $r$. For a quantitative comparison with Whitham modulation theory we also report, as vertical dashed lines superimposed on the measured data, the delays  $T_j= \sqrt{k'' \gamma P_R} ~\tau_j ~L$, $j=1,2,3,4$ corresponding to the edges of the RWs in Eq. (\ref{RW}) and the DSW in Eq. (\ref{DSW}), respectively. We remark that: (i) the experimental traces (left column) show a very good agreement with both simulations based on the NLSE (right column), and the predicted delays $T_j$ from modulation theory; (ii) the duration of the plateau connecting the DSW and the elevating RW significantly shrinks as the ratio $r$ decreases (from top to bottom in Fig. \ref{fig3}), again in quantitative agreement with modulation theory which predicts that $|\tau_2-\tau_3|$ reduces when $r$ decreases;
(iii) for relatively large ratios $r$ the DSW never touches zero and the edge $\tau_2$ of the DSW (where $m=1$) is constituted by a gray soliton \cite{Hoefer06}, as shown in Fig. \ref{fig3}(a1-a2). 
However, by decreasing $r$, we observe the onset of cavitation when the soliton at the trailing edge of the DSW becomes black (see Fig. \ref{fig3} (b1,b2), $r = 0.11$). The observed threshold value $r=0.11$ is in excellent agreement with the theoretical prediction [Eq. (\ref{th})]. While the black soliton possess a zero velocity (with respect to its background), the DSW edge maintains a finite velocity due to the chirped background (velocity $u_i$). Also associated with such black soliton we expect a phase jump of $\pi$ (see Fig. S5(b) for more details) which, however, cannot be measured with our set-up. At even lower ratios $r$, the vacuum point shifts towards the left and the DSW envelope becomes non-monotone.

\begin{center}
\begin{figure}[t!]
\includegraphics[width=8.3cm]{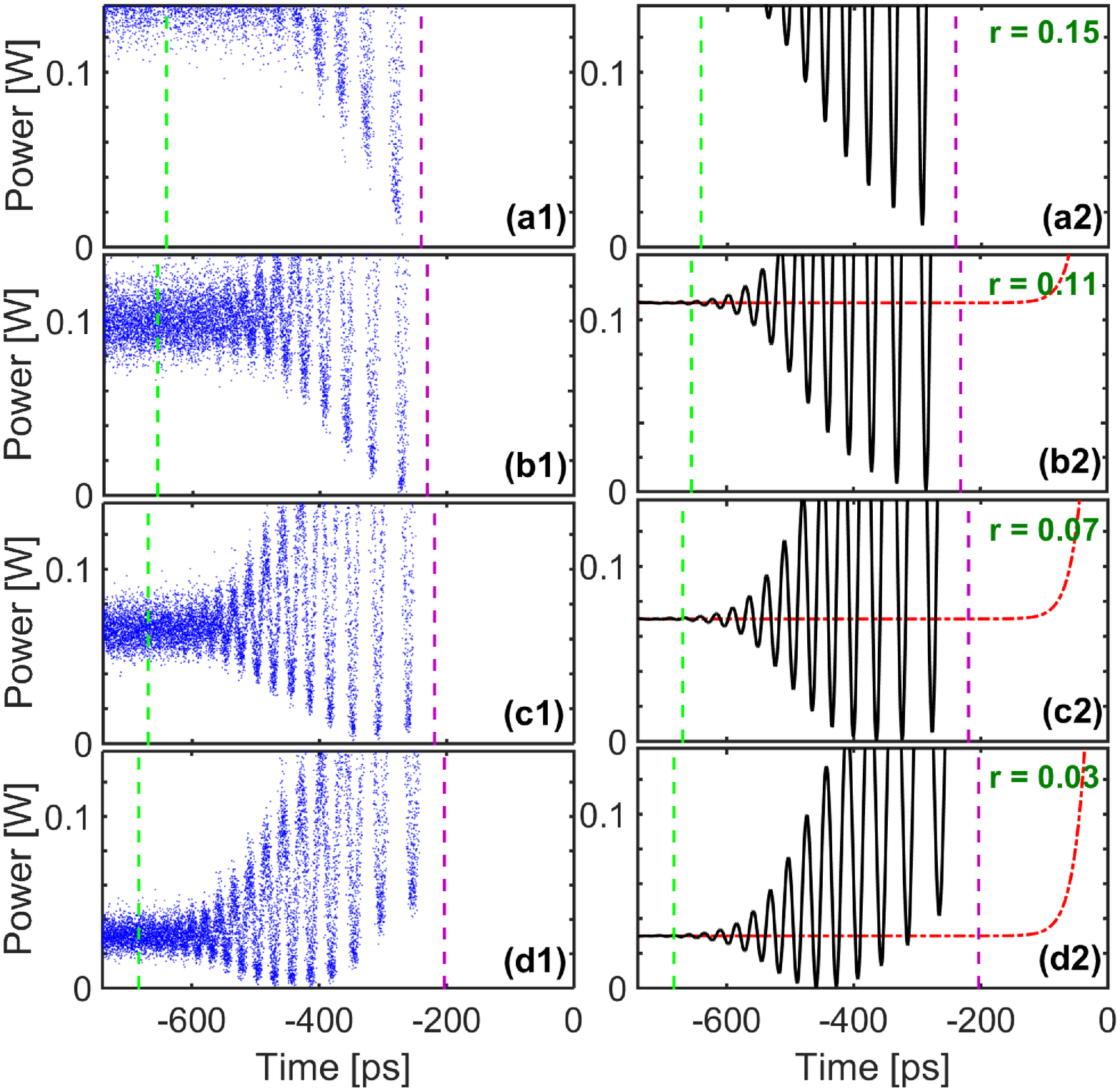}
\includegraphics[width=8.3cm]{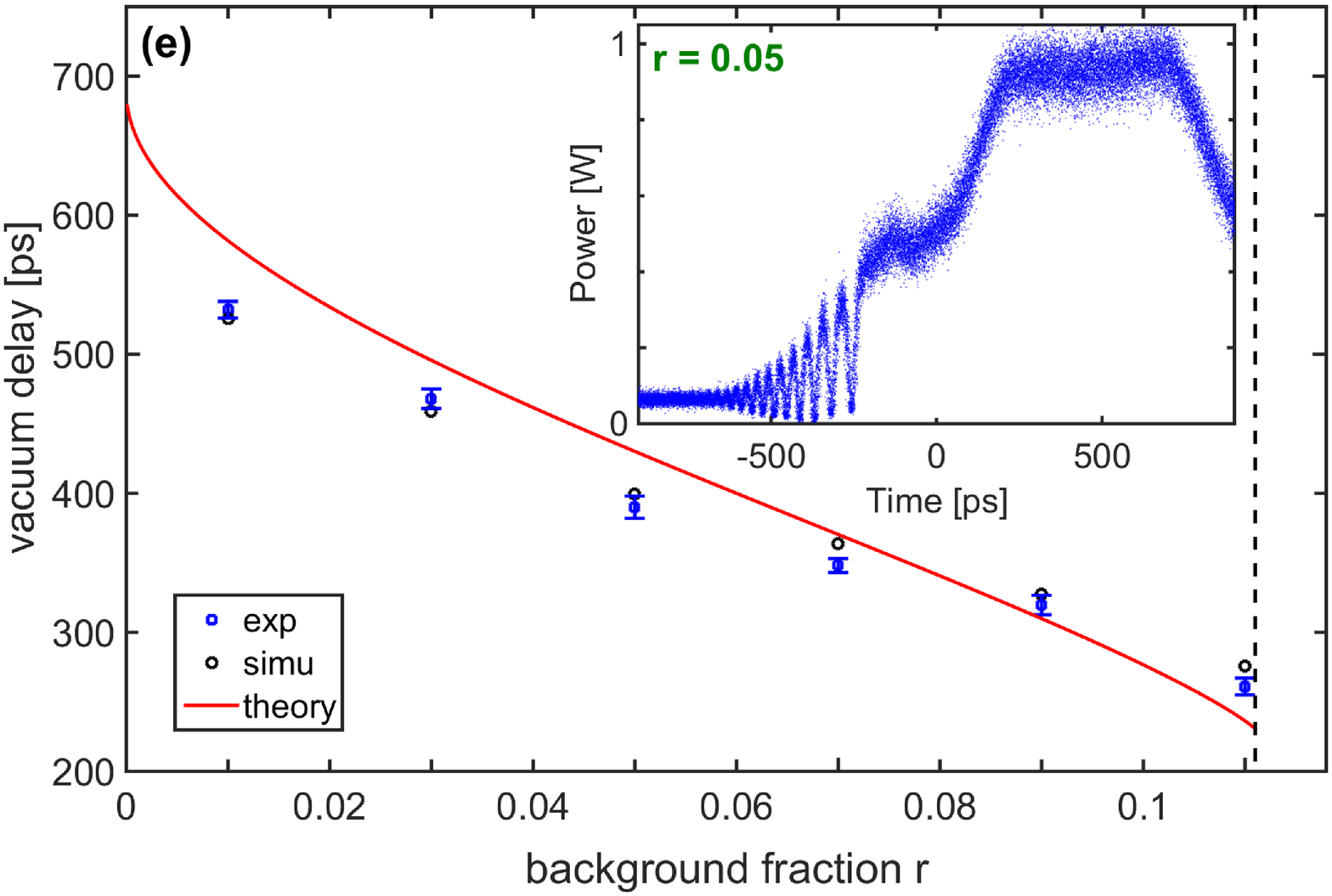}
\caption{
(a-d) As in Fig. \ref{fig3}, showing a zoom around the spontaneously cavitating state (vacuum point) of the DSW, for $P_R=1$ W and different fractions $r=P_L/P_R$: (a1,a2) $0.15$; (b1,b2) $0.11$; (c1,c2) $0.07$; (d1,d2) $0.03$.
(e) Measured delay of the cavitating state vs. $r$, compared with prediction from Eq. (\ref{vacuum}) and numerical simulations of the NLSE. 
The dashed vertical line stands for the threshold in Eq. (\ref{th}). Inset: overall RW-DSW dynamics of the case $r$ = 0.05.}
\label{fig4}
\end{figure}
\end{center}

While the results in Fig. \ref{fig3} already show the crossover to the cavitating state, a detailed quantitative study of this regime requires to operate with a DSW possessing larger extension and shorter average period. To this end, we operate at the maximum available power in our set-up, $P_R=1$ W. This allows us to observe a DSW exhibiting several oscillations with an average period of $\sim 30$ ps, spanning a range that exceeds $400$ ps, as shown in the inset of Fig. \ref{fig4}(e), for $P_L=50$ mW or $r=0.05$.
Importantly,  in this regime the delay of the cavitating state (zero power) can be accurately identified within the DSW. Figures \ref{fig4}(a-d) display a zoom over the bottom part of the DSW in order to show how the cavitating state moves when the ratio $r$ is varied across the threshold. As shown in Fig. \ref{fig4}(a1-a2), when $r=0.15$, the DSW still exhibits monotone envelopes featuring a gray soliton edge with non-vanishing dip.
However, Fig. \ref{fig4}(b1-b2) show very clearly the onset of cavitation (black soliton edge) at the threshold $r=0.11$, in agreement with Eq. (\ref{th}). Decreasing further $r$ (i.e., for higher dam heights) leads the cavitating state to acquire increasingly negative delays, shifting progressively towards the linear edge of the DSW, as clear from Fig. \ref{fig4}(c1-c2) for $r=0.07$ and Fig. \ref{fig4}(d1-d2) for $r=0.03$. For all cases the numerical simulations are still in good agreement with the measured profiles. We summarise in Fig. \ref{fig4}(e) the delays of the vacuum state extracted from the measured temporal traces for different ratios $r$ (see also \cite{supplementary}) in the range $ 0.01 \le r \le r_{th}$ (below $r \sim 0.01$, the residual noise makes impossible to resolve the delay of the vacuum which is quite close to the linear edge). The data are contrasted with the theoretical prediction from Eq. (\ref{vacuum}), showing a satisfactory agreement in the whole range.
We ascribe the discrepancies to the finite rise-time of the step and to the asymptotic character of Whitham theory, which is expected to become more accurate as the propagation length increases and/or the dispersion decreases \cite{Grava}.

In summary, we have reported the fluid behavior of light in a dispersive dam-break experiment, revealing a transition to cavitation in close quantitative agreement with the predictions of modulation theory.
Such behavior is expected to be universal for systems ruled by the NLSE, while qualitatively differing from other dispersive breaking scenarios observed in fluids \cite{dispdambreakwater}.
Our platform could be further used to explore other critical behaviors in the general dispersive Riemann problem \cite{El_PhysD95,bk06}, including the focusing case \cite{focusing}.     

 
\begin{acknowledgments}
The present research was supported by IRCICA (USR 3380 CNRS), by the Agence Nationale de la Recherche in the framework of the Labex CEMPI (ANR-11-LABX-0007-01), Equipex FLUX (ANR-11-EQPX-0017), by the projects NoAWE (ANR-14-ACHN-0014), TOPWAVE (ANR-13-JS04-0004), and the Ministry of Higher Education and Research, Hauts de France council and European Regional Development Fund (ERDF) through the Contrat de Projets Etat-Region (CPER Photonics for Society P4S). S.T. acknowledges also the grant PRIN 2012BFNWZ2. The authors are grateful to L. Bigot, E. Andresen and IRCICA-TEKTRONIX European Optical and Wireless Innovation Laboratory for technical support about the electronic devices.
\end{acknowledgments}

\cleardoublepage

\setcounter{equation}{0}
\setcounter{figure}{0}
\renewcommand{\thefigure}{S\arabic{figure}}
\renewcommand{\theequation}{S\arabic{equation}}

\section{Supplemental material}

\begin{figure}[ht]
\includegraphics[width=8.1cm]{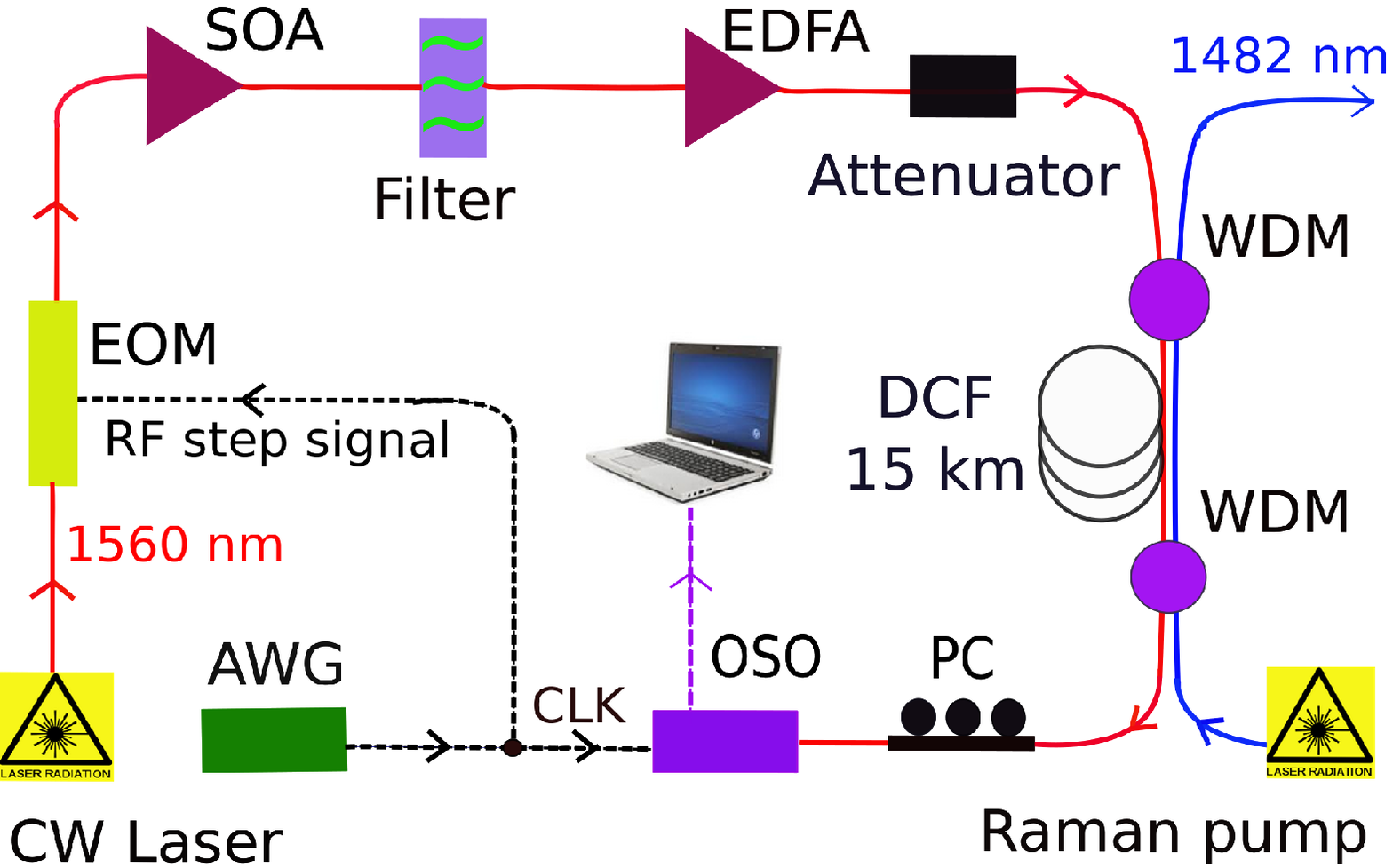}
\caption{
Experimental setup: EOM electro-optic modulator; EDFA, erbium-doped fiber amplifier; AWG, Arbitrary wave generator; OSO, optical sampling oscilloscope; SOA, semiconductor optical amplifier; WDM, Wavelength Division Multiplexing; DCF, dispersion compensating fiber; PC, Polarization controller.}
\label{fig0}
\end{figure}

\begin{center} {\bf 1. Details on the experimental setup} \end{center}
In this section, we provide more details on the experimental setup sketched in Fig. \ref{fig0}.

The light source is a continuous-wave laser diode centered at 1560 nm which delivers 5 mW of power. The two steps in power are generated by means of an electro-optic modulator (EOM, NIR-MX-LN series, 20GHz of bandwidth, Photline) driven by an arbitrary waveform generator (AWG7000, 50 GHz of bandwidth, Tektronix). The signal is pre-amplified in a semiconductor optical amplifier, the spontaneous emission in excess is removed by a spectral filter of 1 nm bandwidth (full width half maximum), and further amplified with an erbium-doped fiber amplifier (EDFA). The power can be tuned  by means of a variable attenuator before being launched inside the optical fiber. Temporal traces are recorded with an optical sampling oscilloscope with a 1.6 ps resolution (Eyechecker, Alnair). Overall, we obtain a clean step-like optical waveform with controllable power levels and durations, as well as a very short rise time ($10-90\%$ rise time is $\sim 50$ ps), which is suitable to observe the dam-break phenomenon. The repetition rate was fixed to 78 MHz that corresponds to a good trade off between a large enough value required by the sampling oscilloscope to get stable behavior and a sufficiently low value to obtain high peak power by using the EDFA. 
\begin{figure}[ht]
\includegraphics[width=8.1cm]{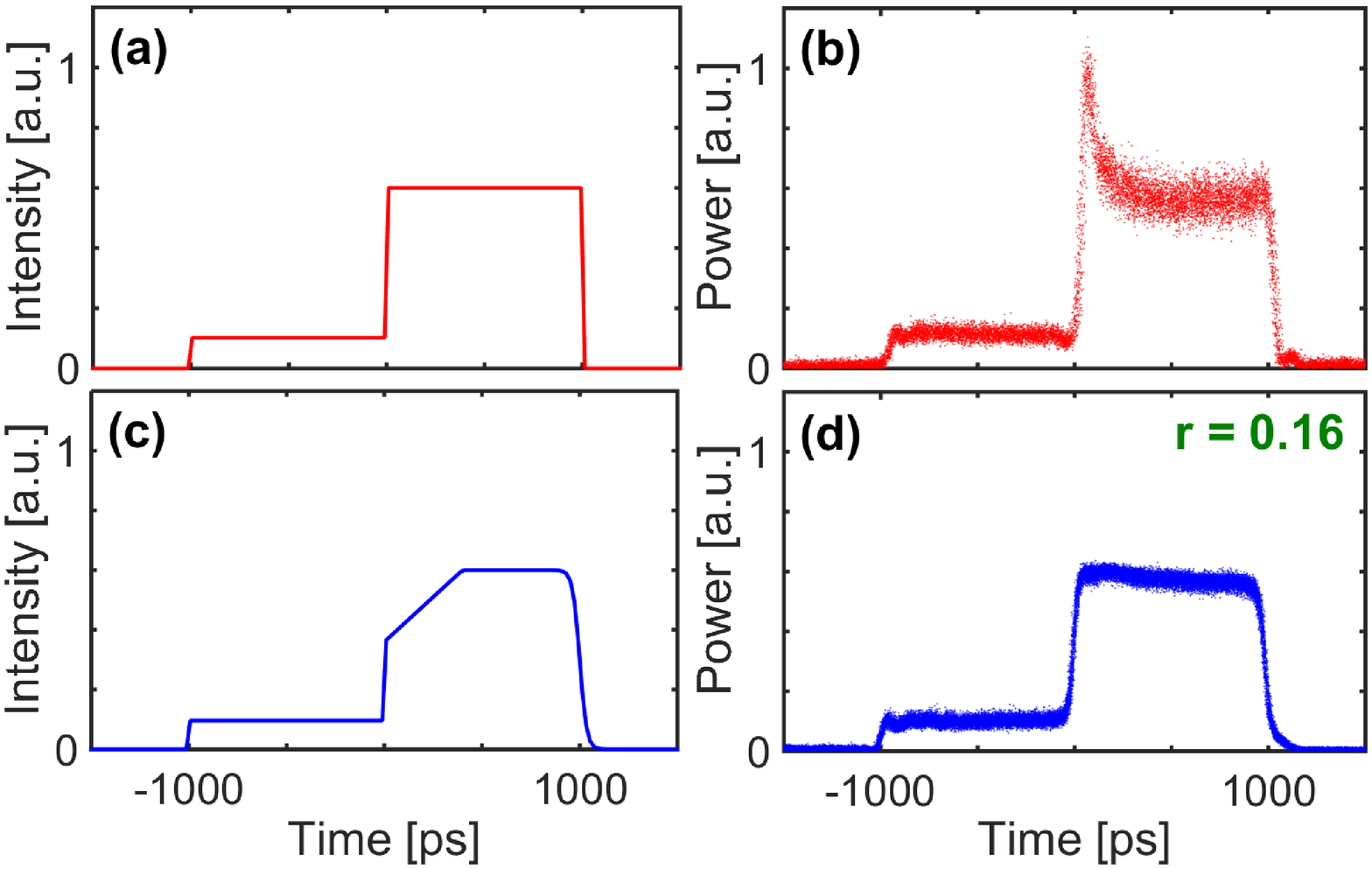}
\caption{
Essential pre-compensation to obtain a clean step-like initial condition as in Fig. 2(a). Input programmable waveforms from the AWG (left column) and the corresponding generated optical step-like input (right column) to be launched in the fiber. The case with (without) pre-compensation of distortions are the blue (red) traces.}
\label{precompensation}
\end{figure}
The distortions induced by the EOM and the nonlinear response of the amplifiers are pre-compensated by finely adjusting the shape of the electrical signal delivered by the AWG. As can be seen in Fig. S1, in order to obtain the clean step-like signal at the fiber input (as in Fig. 2(a) of the letter, and Fig. \ref{precompensation}(d)), it was necessary to drive the modulator with the signal shown in Fig. \ref{precompensation}(c) to avoid the unexpected overshoots at $t \sim 30,1070$ ps shown in Fig. \ref{precompensation}(b), which result from driving the AWG with the nearly ideal signal in Fig. \ref{precompensation}(a).

\begin{figure}[t!]
\includegraphics[width=8.1cm]{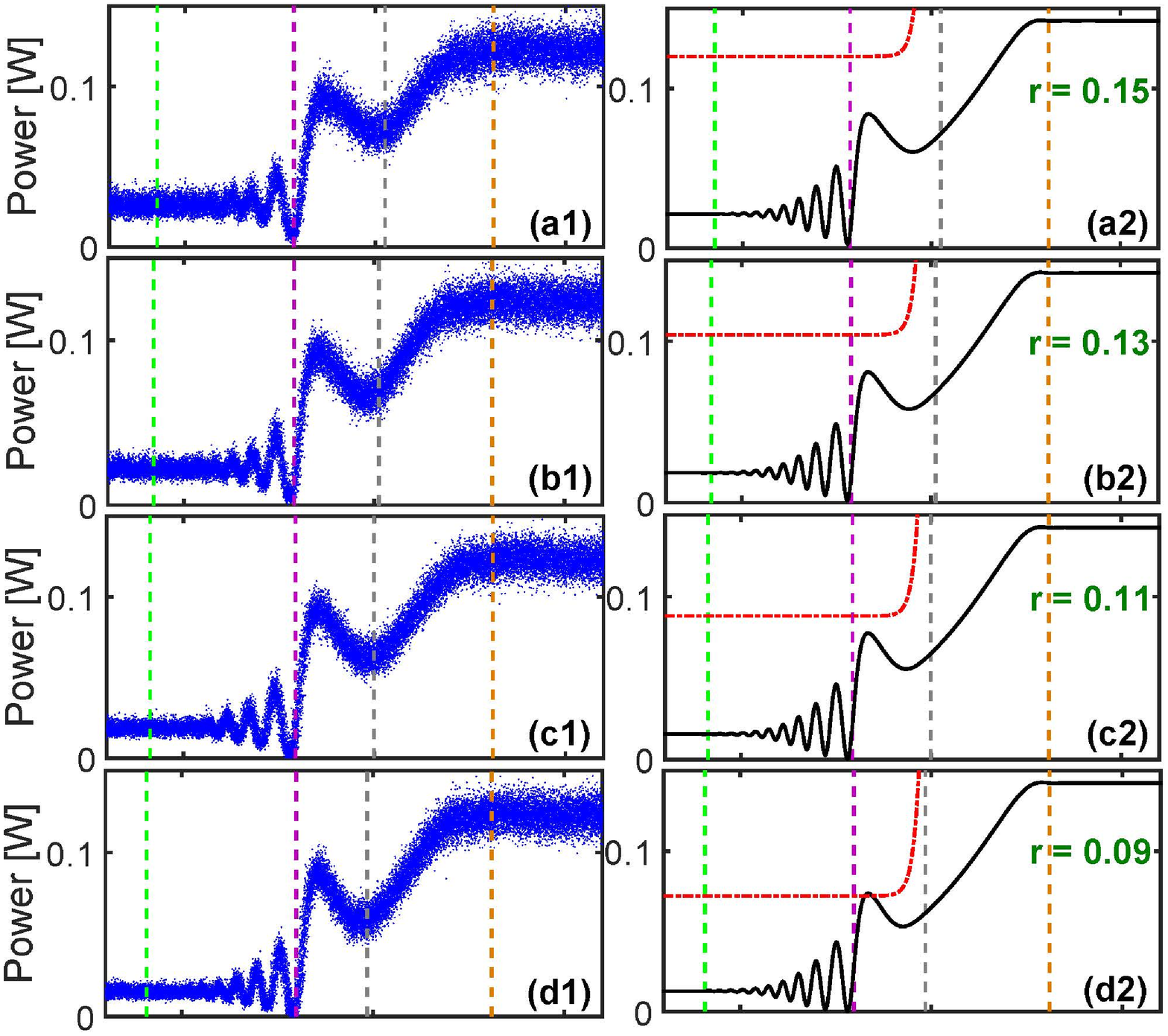}
\includegraphics[width=8.2cm]{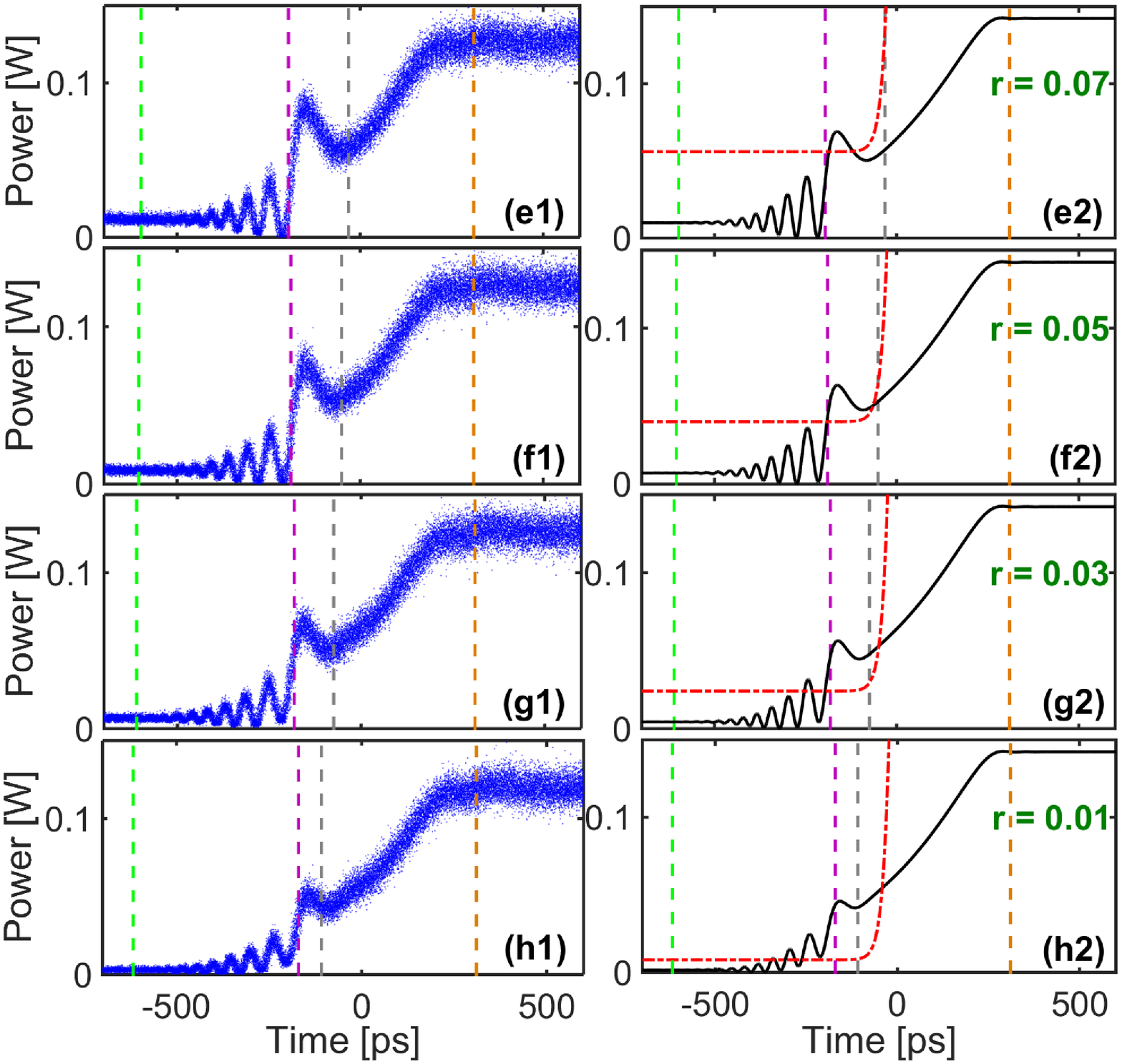}
\caption{Temporal traces at the fiber output in the uncompensated scheme (fiber loss of 0.5 dB/km), for $P_R = 0.8$ W and different background fractions $r=P_L/P_R$:
(a1,a2) r=0.15; (b1,b2) 0.13; (c1,c2) 0.11; (d1,d2) 0.09; (e1,e2) 0.07; (f1,f2) 0.05; (g1,g2) 0.03; (h1,h2) 0.01; Left column: experimental results. Right column: numerical simulations based on the NLSE with linear losses. The vertical dashed lines give the predicted delays (gray and orange lines, edge of RW from Eqs. (4) of the letter; magenta and green lines, edges of DSW from Eqs. (5) of the letter), respectively. Dashed-dot red lines is the numerical step-like input.}
\label{No_Compensation}
\end{figure}
\begin{figure}[t!]
\includegraphics[width=8.1cm]{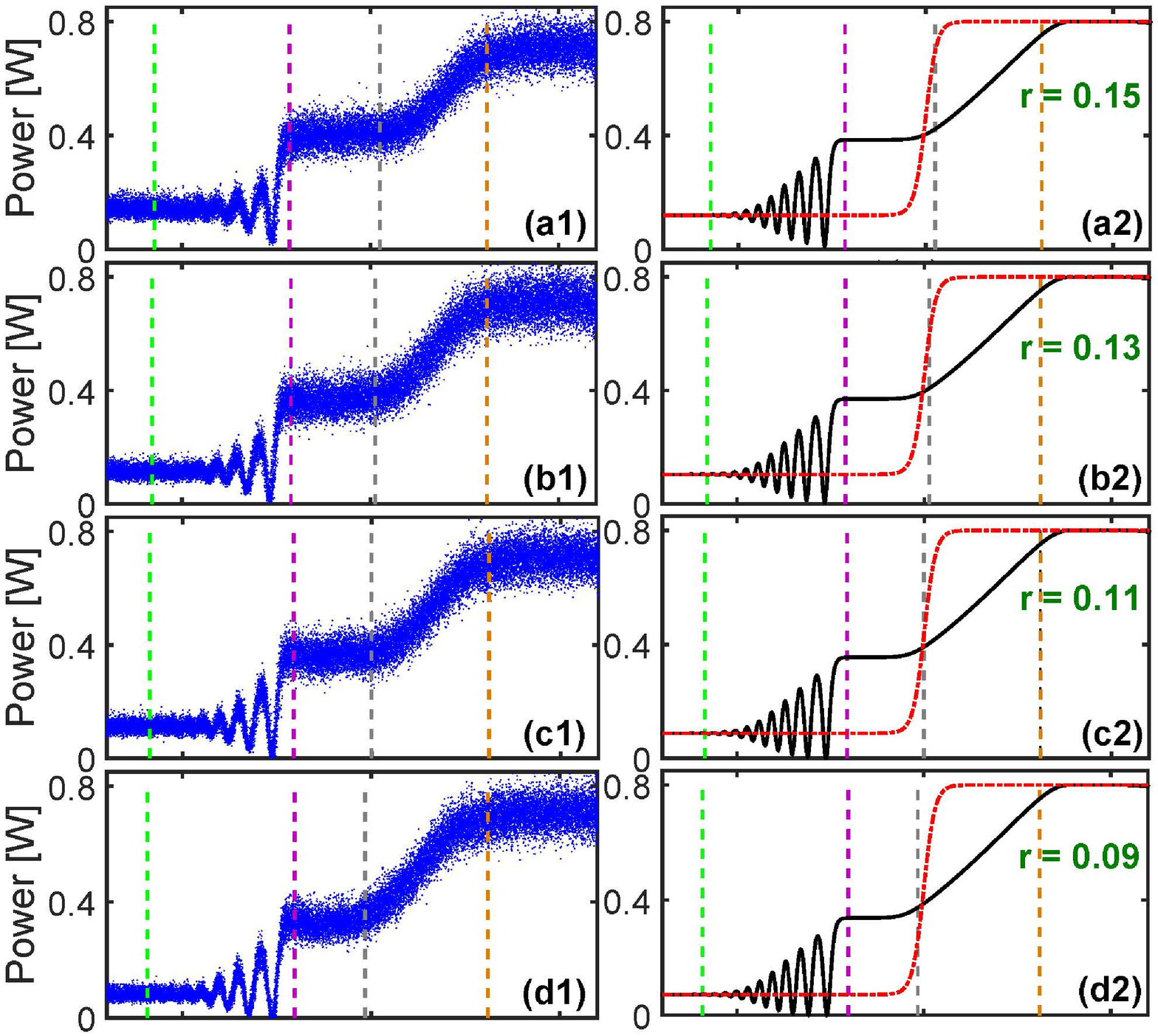}
\includegraphics[width=8.1cm]{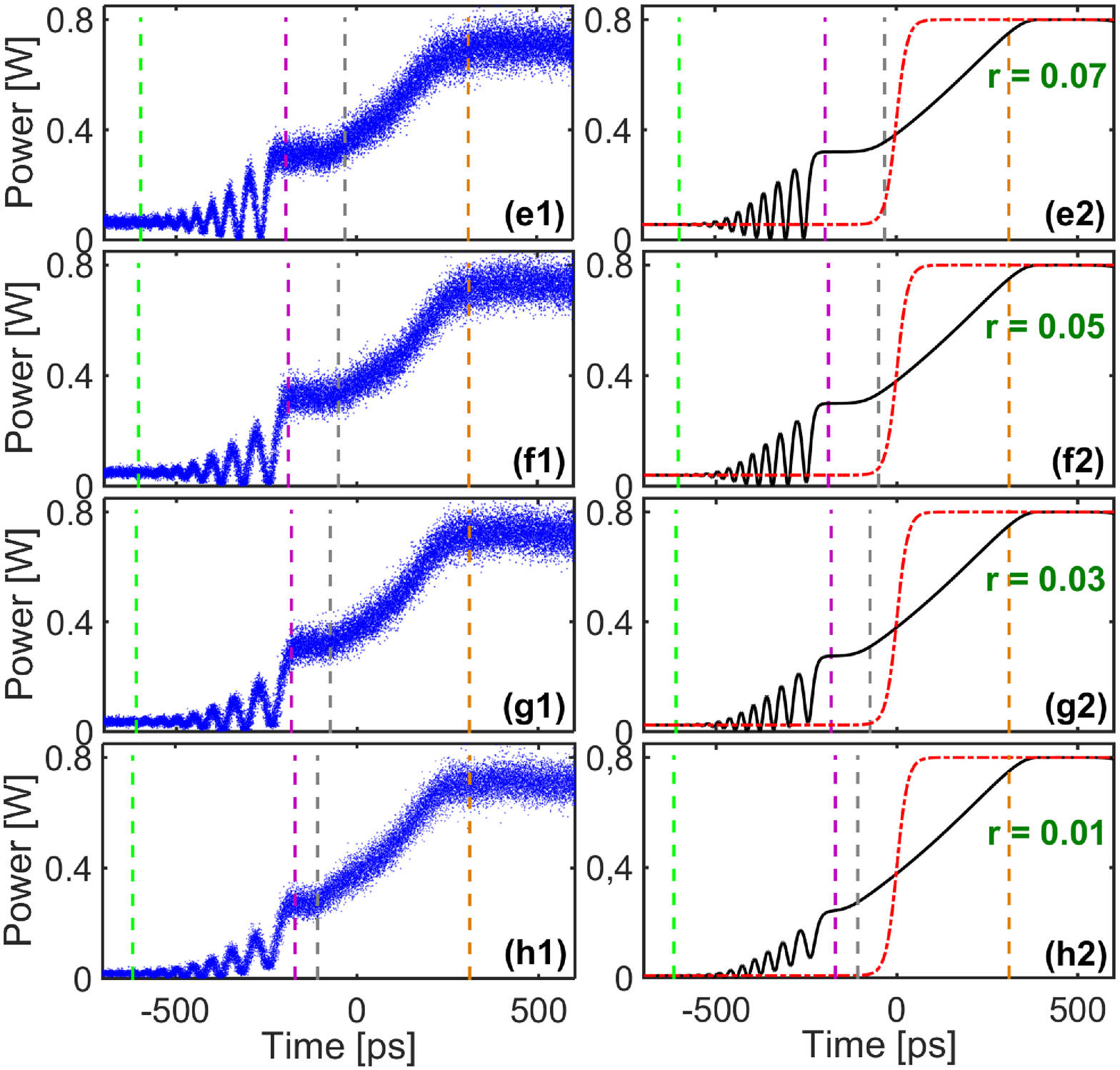}
\caption{Extended version of Fig. 3 in the letter. Here the panels have one to one correspondence with those in Fig. \ref{No_Compensation}, except that the data are collected with activated Raman amplification, which allows to effectively compensate for the linear loss in the fiber.}
\label{With_Compensation}
\end{figure}
An other crucial feature of the set-up is the loss compensation scheme. 
Indeed, while the losses in the optical fiber are rather low (0.5 dB/km), the long fiber length (15 km) leads to 7.5 dB ($\sim 80\%$) of total attenuation. This strongly affects the dynamics of the process as evident by contrasting Fig. \ref{No_Compensation}(a-h) with Fig. \ref{With_Compensation}(a-h) that depict the regimes with and without losses for different values of $r=P_L/P_R$ and the same input power $P_R=0.8$ W.  The losses, besides substantially damping the output power (compare vertical scales in Figs. \ref{No_Compensation} and \ref{With_Compensation}), determine a strong deformation of the characteristic plateau that separates the RW and the DSW, which is an essential trait of the fluid-like dam-break scenario. In order to compensate this deviation from the ideal fluid behavior, we compensate the fiber losses by taking advantage of Raman amplification as employed in telecommunication systems \cite{AgrawalRaman}. The aim is to get an effectively transparent fiber.  
We used a Raman pump of 2 W maximum power centered at 1482 nm ($\sim 13$ THz detuning from the signal, which guarantees a large Raman gain \cite{AgrawalRaman}), which is launched and extracted by means of multiplexers at the fiber input and output. The pump counter-propagates with respect to the signal in order to minimise the pump to signal relative intensity noise transfer \cite{AgrawalRaman}. We carefully adjusted the Raman pump power by looking at the output traces until we achieve both a flat plateau between the DSW and the RW and a peak power plateau that equals the input level. The validity of the scheme is clear by comparing the left columns in Fig. \ref{No_Compensation} (Raman pump off) with those in Fig. \ref{With_Compensation}
(Raman pump on, loss compensated), and from the perfect agreement that we obtain with numerical simulations (right columns) based on the NLSE with included linear loss in Fig. \ref{No_Compensation} (a2-h2) and the pure (conservative) NLSE in Fig. \ref{With_Compensation} (a2-h2). To summarise, our fiber system accurately mimics the dispersive hydrodynamics described by the conservative NLSE.
\begin{center} {\bf 2. Additional experimental results on the self-cavitation transition} \end{center}
\begin{figure}[h!]
\includegraphics[width=8cm]{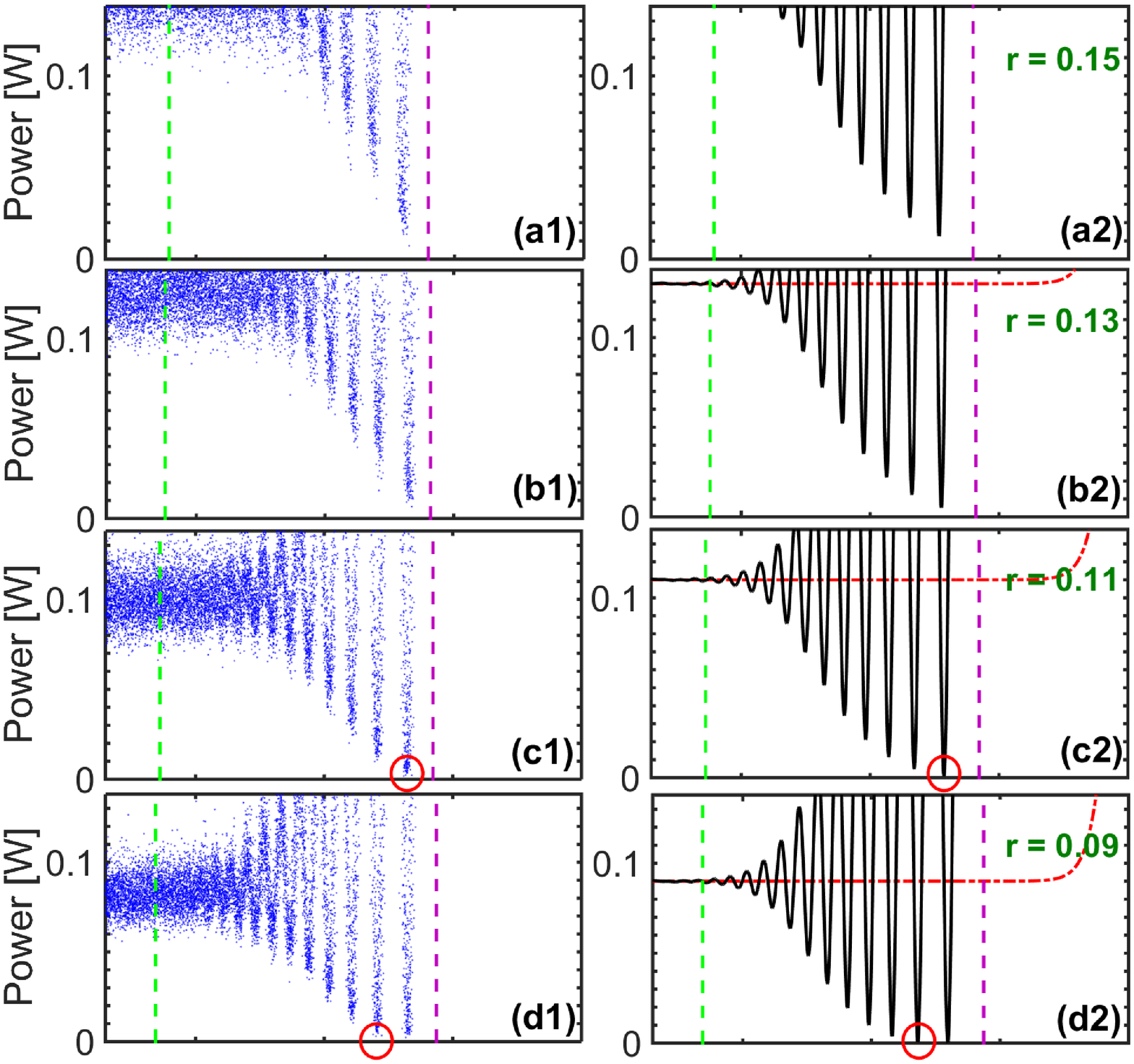}
\includegraphics[width=8.1cm]{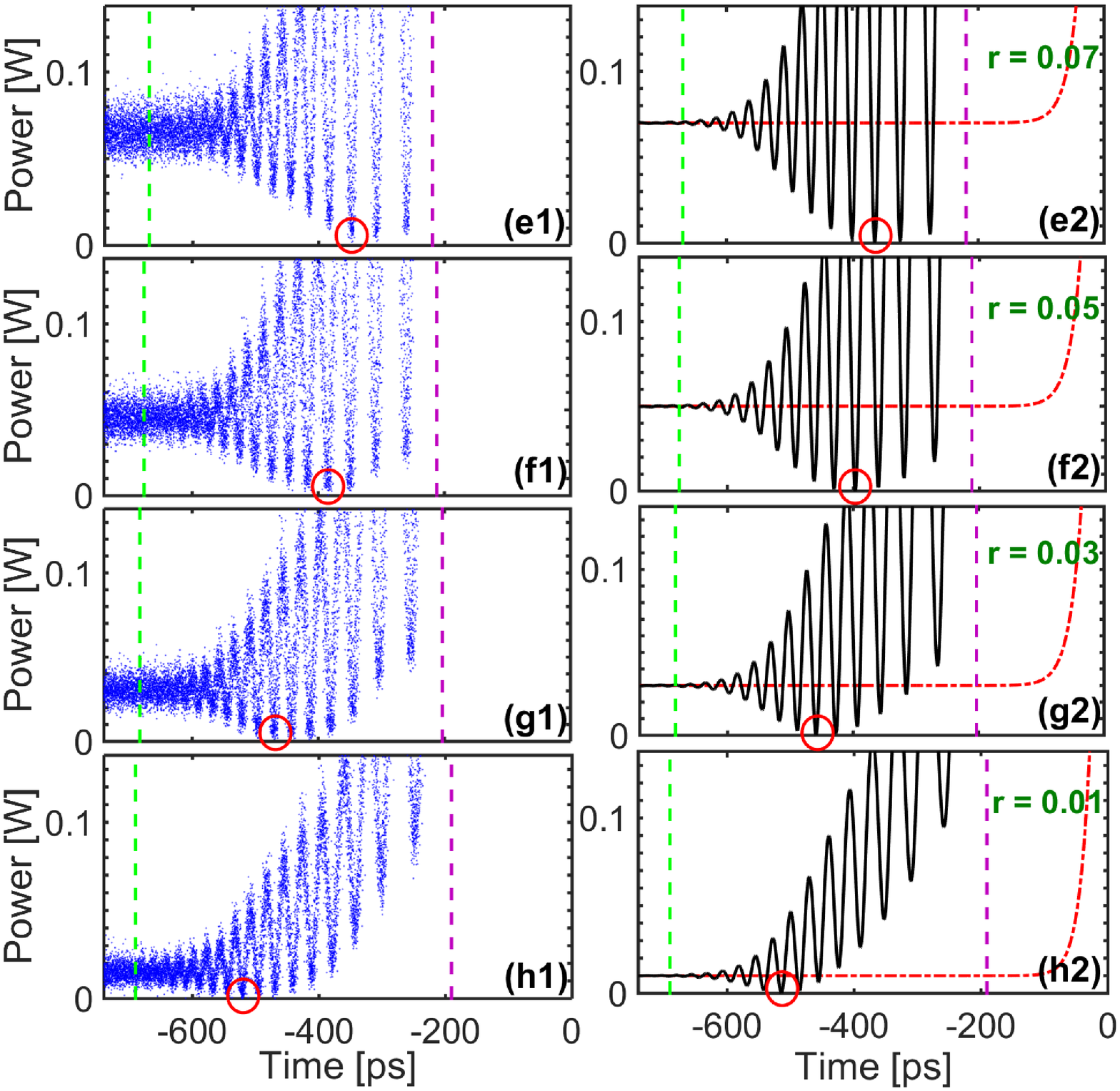}
\caption{Additional results on the DSW dynamics around the cavitating state (cf. Fig. 4 in the paper), for different ratio $r$: (a) 0.15; (b) 0.13; (c) 0.11 (cavitation threshold); (d) 0.09; (e) 0.07; (f) 0.05; (g) 0.03; (h) 0.01. Left column: experimental results; Right column: numerical simulations (solid-black line is the output, dashed red line is the input).}
\label{Vacuum_points}
\end{figure}

We report in Fig. \ref{Vacuum_points} the full sequence of close-up measurements (with corresponding simulations) of the DSW bottom for different values of $r=P_L/P_R$ (Fig. 4 in the letter contains a subset of such measurements). Above threshold $r >r_{th}=1/9$ the DSW envelope is monotone and no vacuum appears (Fig. \ref{Vacuum_points}(a-b)). Conversely, below threshold, a self-cavitating state (vacuum) $r \le r_{th}$ appears, as highlighted by red circles in Fig. \ref{Vacuum_points}(c-e).
The large extension of the DSW due to the high power regime ($P_R=1$ W) allows us to observe with extremely good resolution the shift of the vacuum towards the leading (linear) edge of the DSW as $r$ is reduced. A very good agreement is also achieved with numerical simulations. 

\begin{center} {\bf 3. Numerics: further details} \end{center}
All the simulations based on the NLSE are made by means of the split-step method. The elevating step from $P_L$ to $P_R$ is modelled by the input field $E(Z=0,T)=[P_L+(P_R-P_L)(1+\tanh(T/T_r))/2]^{1/2}$, where $T_r=25-30$ ps arises from the fit with the optical input trace at any run, and is due to the rise-time of the modulator. Similarly, for the decreasing step on the trailing (``dry-bed") edge we use $E(Z=0,T)=[P_R(1-\tanh(T/T_{f}))/2]^{1/2}$ with $T_{f}=45$ ps due to slightly longer falling time of the modulator. Obvious combinations of the steps allow for modelling the full waveform as in Fig. \ref{precompensation}(d).

\begin{center}
\begin{figure}[h]
\includegraphics[width=8.8cm]{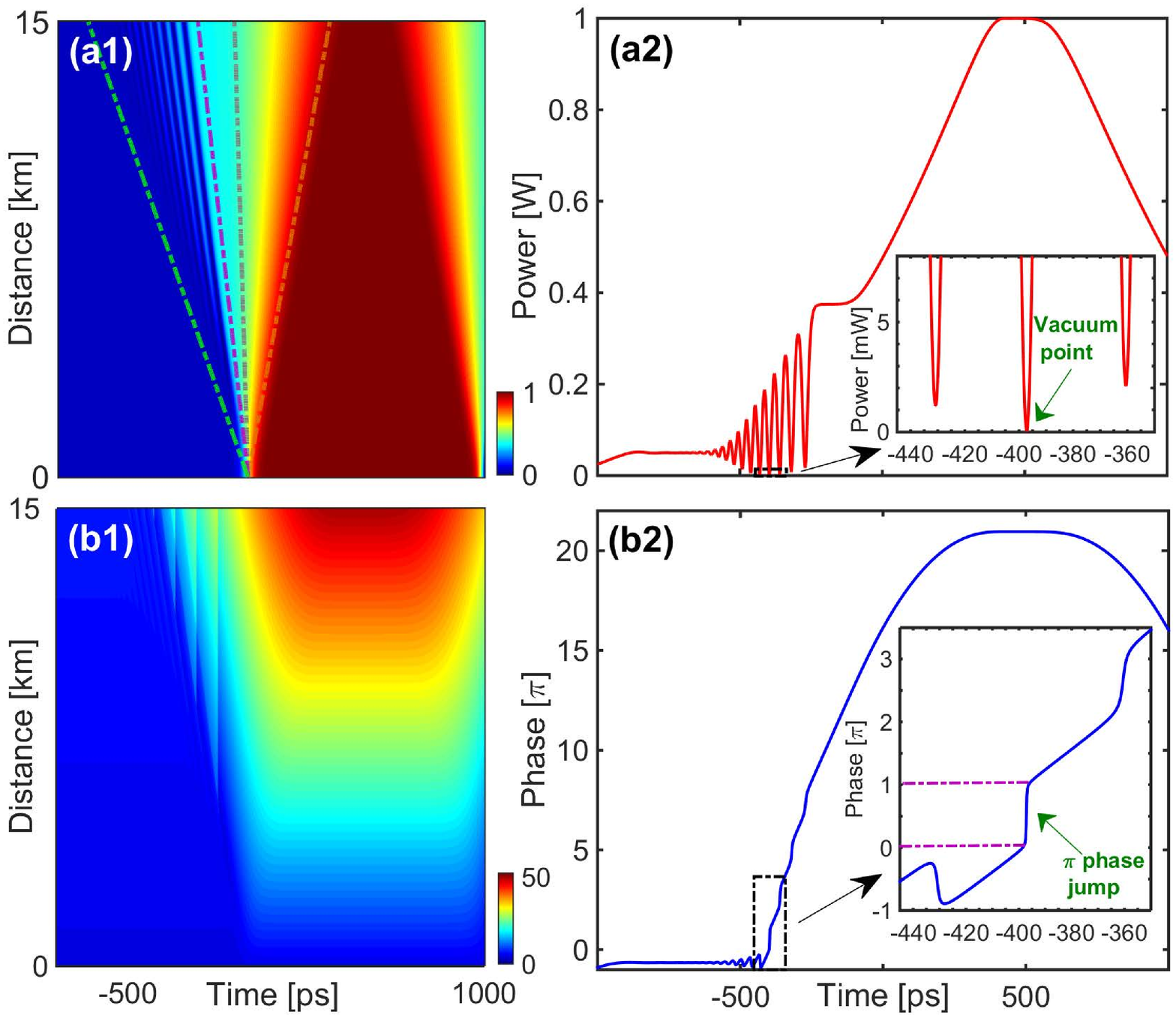}
\includegraphics[width=8.8cm]{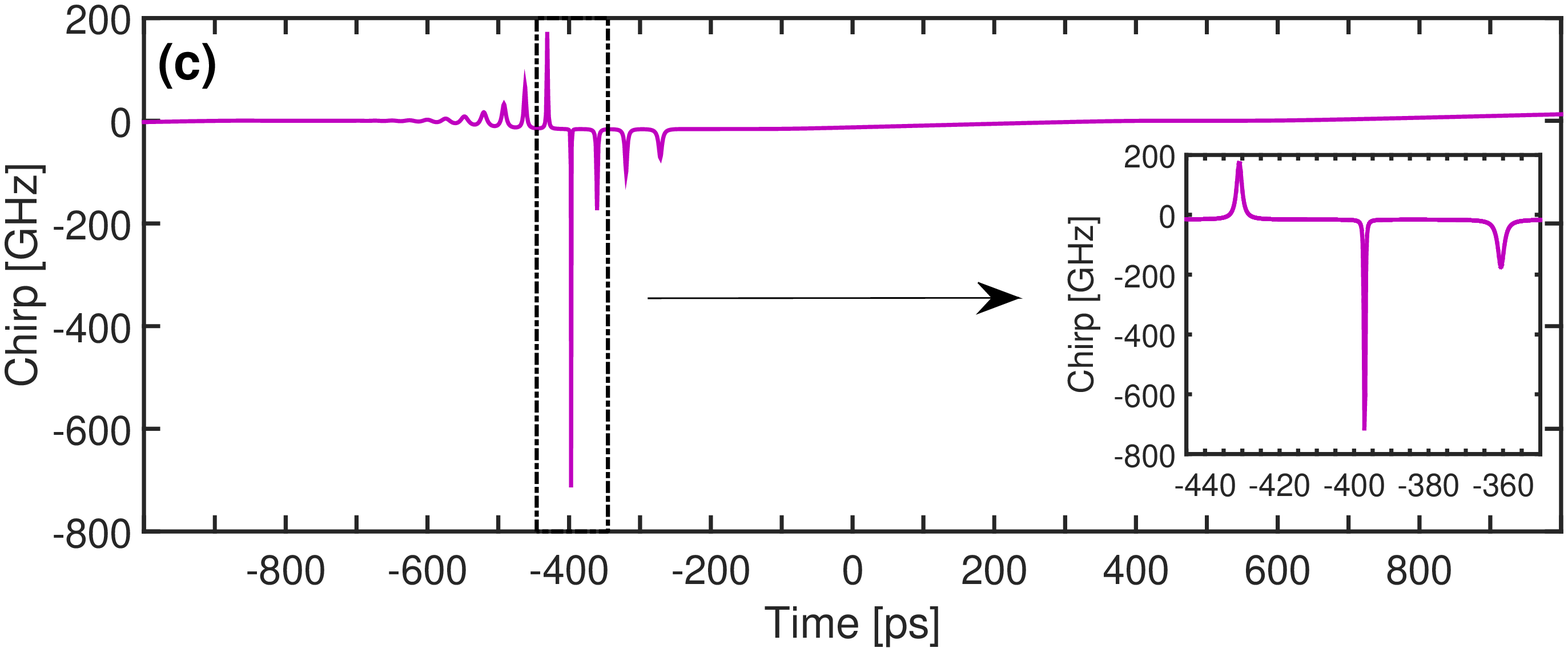}
\caption{
Numerical simulation based on the NLSE for the case shown in Fig. \ref{Vacuum_points}(f2). Level color plot of power (a1) and phase (b1) evolution along the fiber. Power (a2) and phase (b2) profiles at the fiber output. Dashed-dot lines in (a1) correspond to the analytical predictions for the DSW and RW edges, respectively [Eqs. (4-5) in the letter]. The insets of (a2) and (b2) zoom on the DSW bottom to emphasise the self-cavitating state (null power) and the corresponding phase jump of $\pi$. (c) output frequency chirp $u$, with close-up showing the change of sign of the chirp.
}
\label{phase_intensity}
\end{figure}
\end{center}

We display in Fig. \ref{phase_intensity}(a) and (b) an example of the full evolution of the intensity and phase along the fiber, for $r=P_L/P_R = 0.05$, which corresponds to Fig. \ref{Vacuum_points}(f1,f2). The dot-dashed oblique lines in Fig. \ref{phase_intensity}(a1) stand for the edges of the DSW and RW from Eqs. (4-5) of the letter. From Fig. \ref{phase_intensity}(a1) and Fig. \ref{phase_intensity}(a2) reporting the output power profile, we clearly see the expanding plateau between the DSW and the RW. We also notice that the top of the pulse (peak power of 1W) is eroded by the two RWs (the left and right RW are due to the ``wet-bed" and ``dry-bed" dynamics on the input leading and trailing edges, respectively), without any appreciable interactions between them (this witnesses that the two scenarios do not mix up over the finite fiber length). A close-up of the bottom of the shock fan is shown in the inset of Fig. \ref{phase_intensity}(a2). Around the vacuum point ($t \simeq -398$ ps), the numerics reveals a $\pi$ phase jump as shown in the inset of Fig. \ref{phase_intensity}(b2), consistently with the local nature of black solitons. Across the vacuum the chirp (phase derivative) $u$ exhibits strong peaks of different sign (see Fig. \ref{phase_intensity}(c)). Unfortunately such fast phase features are not measurable with the present set-up and will be investigated in the future by means of phase reconstruction techniques.
        
\begin{center} {\bf 4. The velocities: from hydrodynamic approximation to modulation theory} \end{center}

Since the shallow water equations (SWEs) are a two-dimensional vector conservation law, the dynamics of the general Riemann initial value problem entails the superposition of a wave pair, where each of the wave can be of rarefaction wave (RW) type or classical shock wave (SW) type. A classification into (i) RW-RW; (ii) SW-SW; (iii) RW-SW, can be made according to the values of the step in $\rho$ and $u$. Such classification can be suitably extended for the dispersive problem ruled by the NLSE \cite{EL1995PhysD,Hoefer2006PRA}. The dam-break Riemann problem corresponds to a step in the variable $\rho(t,z=0)$ with $u(t,z=0)=0$, and leads to a RW-SW pair separated by a plateau with constant $\rho_i, u_i$ . For an increasing step ($\rho_L < \rho_R$), as in the experiment, the SW is left-going, heading towards $t<0$ (whereas, it would be right-going for a decreasing step $\rho_L > \rho_R$).

The first solution to the ``wet-bed" problem ($\rho_L \neq 0$) in the SWEs was reported by Stoker \cite{Stoker}. Essentially one can start from the Rankine-Hugoniot conditions for the SW, and using also the constancy of $u+2\sqrt{\rho}$ across the RW, one can eliminate $u_i$ to achieve an equation to be solved for $\rho_i$. Then, returning to the starting equations, one can specify all the quantities of interest ($u_i$, SW velocity, edges of RW) as a function of the  quiescent states $\rho_R, u_R=0$ and $\rho_L, u_L=0$.\\

\begin{center}
\begin{figure}[ht]
\includegraphics[width=8cm]{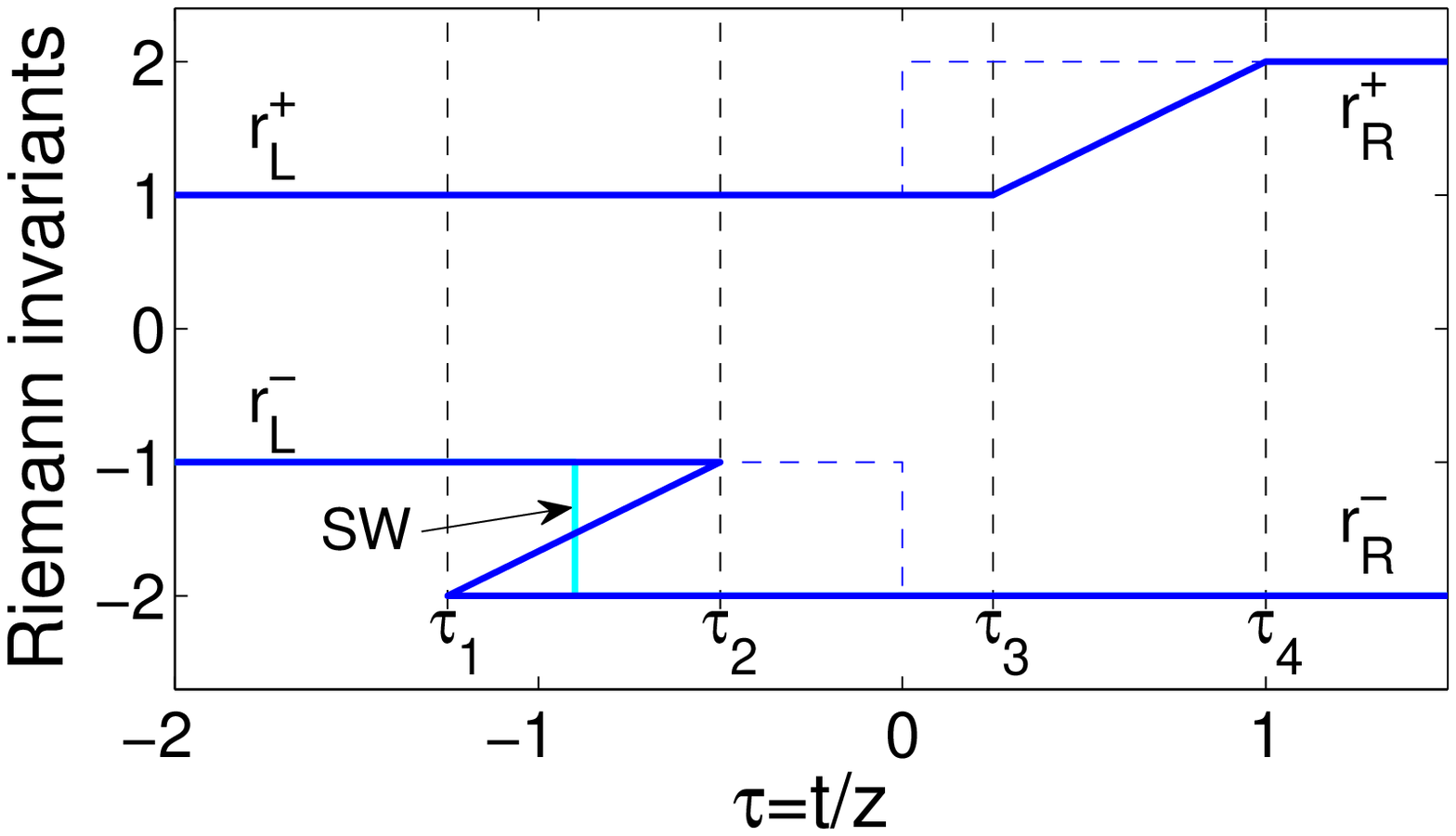}
\caption{
Sketch of the self-similar evolution of the two Riemann invariants $r^{\pm}$ of the SWEs in the dam-break problem. The dashed line steps are the Riemann initial values.
The cyan step is the SW obeying Rankine-Hugoniot conditions.}
\label{S6}
\end{figure}
\end{center}

A slightly different formulation, which is directly suited for the extension to the dispersive case ruled by the full NLSE, amounts to solve in terms of so called simple waves. The starting point are the SWEs cast in diagonal form in terms of Riemann invariants $r^{\pm}(t,z) \equiv u(t,z) \pm 2\sqrt{\rho(t,z)}$,  
\begin{equation} \label{diag}
\partial_z r^{\pm} + v^{\pm} \partial_t r^{\pm}=0,
\end{equation}
where the eigenvelocities are $v^{\pm}=v^{\pm}(r^-,r^+)= (3r_{\pm} + r_{\mp})/4$. 
The dam-break problem with $\rho_L<\rho_R$ implies an increasing [decreasing] step for $r^+(t,0)$ [$r^-(t,0)$], as displayed in Fig. \ref{S6} (steps from $r^+_L=2\sqrt{\rho_L}$ to $r^+_R=2\sqrt{\rho_R}$, and from  $r^-_L=-2\sqrt{\rho_L}$ to $r^-_R=-2\sqrt{\rho_R}$, respectively). For this case, characterised by $r_R^-<r_L^-<r_L^+<r_R^+$, we sketch in Fig. \ref{S6} the simple wave solutions of Eqs. (\ref{diag}), such that one of the Riemann invariant is constant and other one varies in self-similar way, i.e. $r=r(\tau=t/z)$. The solution involves, for $\tau>0$, a constant $r^-=r^-_R$ and a RW for $r^+$ described by $\tau=v^+(r^-_R,r^+)$, whereas for $\tau<0$, $r^+=r^+_L$ is constant, and $\tau=v^-(r^-,r^+_L)$ describes the characteristic overturning (see Fig. \ref{S6}), which allows to introduce the SW through the Rankine-Hugoniot condition (cyan step in Fig. \ref{S6}, see also Fig. 1 of the letter). The upper state of the shock is the plateau in $\rho,u=\rho_i,u_i$, which is found by exploiting continuity in $\tau=0$ to be $u_i=(r^+_L+r^-_R)/2=\sqrt{\rho_L}-\sqrt{\rho_R}$ and $\rho_i=[(r^+_L-r^-_R)/4]^2=(\sqrt{\rho_L}+\sqrt{\rho_R})^2/4$. The approriate limits of the solution give the relevant values of the self-similar variable $\tau_i$, $i=1,2,3,4$ shown in Fig. \ref{S6}, as follows:
\begin{eqnarray}
\tau_4&=&\frac{3r^+_R+r^-_R}{4}=\sqrt{\rho_R} \label{RW1};\\
\tau_3&=&\frac{3r^+_L+r^-_R}{4}=\frac{3\sqrt{\rho_L}-\sqrt{\rho_R}}{2} \label{RW2};\\
\tau_2&=&\frac{3r^-_L+r^+_L}{4}=-\sqrt{\rho_L};\label{SW1} \\
\tau_1&=&\frac{3r^-_R+r^+_L}{4}=\frac{\sqrt{\rho_L}-3\sqrt{\rho_R}}{2}, \label{SW2}
\end{eqnarray}
where we have exploited the fact that $r^{\pm}_{L,R}=\pm 2 \sqrt{\rho_{L,R}}$. 
It is important to emphasise that $\tau_4$ and $\tau_3$ in Eqs. (\ref{RW1}-\ref{RW2}) represent the edge velocities of the RW, which hold valid, due to the RW smoothness, also when dispersive corrections are accounted for.
Conversely, when dispersive correction are applied to the SW, one needs to calculate the edge velocities by averaging over the DSW oscillation as explained below.
We anticipate that this gives a result markedly different from the expression of $\tau_2$ and $\tau_1$ in Eqs. (\ref{SW1}-\ref{SW2}).

In the presence of an undular shock or DSW, the modulation theory based on Whitham averaging can be exploited to describe the DSW as a modulated cn-oidal wave with parameters that varies slowly with respect to its natural periodicity.
The averaged Whitham equations describe the slow variation of the parameters and allow to compute the velocities of the linear and soliton edge of the DSW.
For the details on the method, we refer the reader to the original literature \cite{pavlov,Gurevich1987JETP,EL1995PhysD} and to the excellent reviews on the topic, in particular to Refs. \cite{Hoefer2006PRA,EL2016PhysD}. Below, we only briefly outline the steps that lead us to obtain Eqs. (4-6) in our letter.  

The Whitham equations for the NLSE can be expressed in the following diagonal form, first derived by Pavlov \cite{pavlov}, by introducing four Riemann invariants $r_i=r_i(t,z)$, $i=1,2,3,4$, ordered in such a way that $r_1 < r_2 < r_3 < r_4$, which are a suitable combination of the parameters of the cn-oidal wave \cite{Gurevich1987JETP,EL1995PhysD,Hoefer2006PRA}
\begin{eqnarray} \label{Weqs}
& &\frac{\partial r_i}{\partial z}+ v_i \frac{\partial r_i}{\partial t} =0;\;\;\;i=1,2,3,4.
\end{eqnarray}
Here the velocities $v_i=v_i(r_1, r_2, r_3, r_4)$ depends on combinations of $\{r_i \}$ and elliptic integrals of first ($K(m)$) and second ($E(m)$) kind. For instance the velocity $v_2=v_2(r_1, r_2, r_3, r_4)$, which will be relevant in the following, reads explicitly as
\begin{equation} \label{v2}
v_2 = V + \frac{1}{2}(r_2-r_1) \left[ 1 - \frac{(r_3-r_1) E(m)}{(r_3-r_2) K(m)} \right]^{-1},
\end{equation}
where $V=\frac{1}{4}(r_1+r_2+r_3+r_4)$ and $m=\frac{(r_4-r_3)(r_2-r_1)}{(r_4-r_2)(r_3-r_1)}$.\\

\begin{center}
\begin{figure}[ht]
\includegraphics[width=8cm]{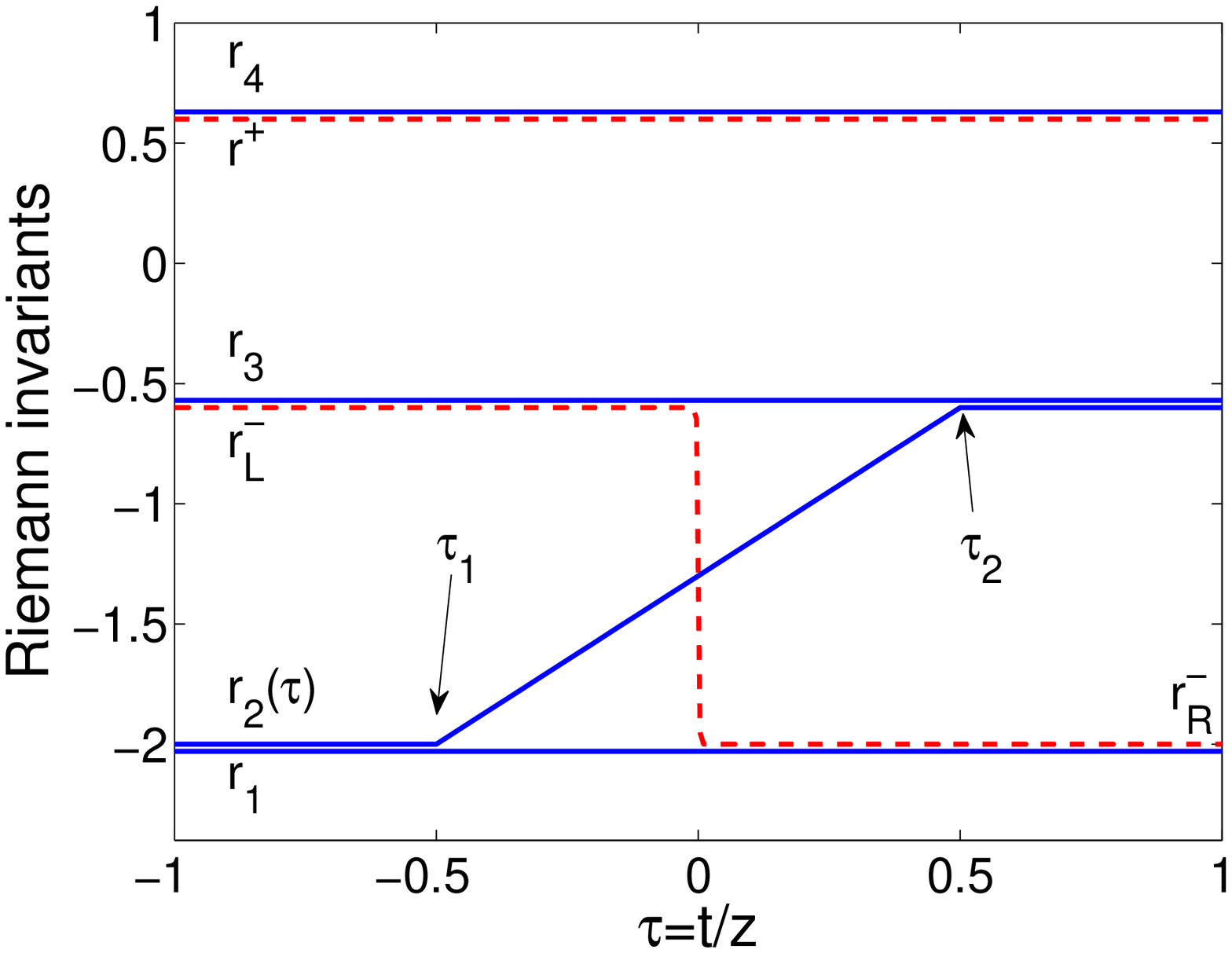}
\caption{
Sketch of the Riemann invariants of Eqs. (\ref{Weqs}), where $r_1, r_3, r_4$ stay constant and $r_2=r_2(\tau=t/z)$ varies smoothly in the range $\tau_1 \le \tau \le \tau_2$ between the values $r_R^-$ and $r_L^-$. $\tau_1$ and $\tau_2$ correspond to the linear and soliton edges of the DSW, respectively. The dashed red lines are the Riemann invariants in the dispersionless limit.}
\label{S7}
\end{figure}
\end{center}

As shown above a left-going classical shock of the SWEs is characterised by a constant $r^+$ and a jump in $r^-$ from $r^-_L$ to $r^-_R$ (with given left and right boundaries $r^-_L,r^-_R$ such that $r^-_R < r^-_L$). The corresponding dispersive regularisation of the shock, i.e. the DSW, can be described by a solution of Eqs. (\ref{Weqs}) where only one of the Riemann invariants varies.
For the left-going DSW, this solution is generated by the initial value $r_{i0}=r_i(t,z=0)$  with $r_{10} =r^-_R$, $r_{30} =r^-_L$, $r_{40} =r^+$, and  a step $r_{20}$ in $t=0$ from $r^-_R$ to $r^-_L$ (obtained from the so-called data regularisation procedure, see \cite{Hoefer2006PRA}). The solution is such that $r_{1}, r_{3}, r_{4}$ remain constant at any $z$, whereas $r_2$ gives rise to a smooth RW (owing to the fact that $r_{20}$ is non-decreasing) that depends on the self-similar variable $\tau=t/z$. Indeed all Whitham equations are formally satisfied when $r_{1,3,4}(t,z)=r_{10,30,40}$ and $r_2(t,z)=r_2(\tau)$, provided the equation $\left( \tau- v_2 \right)~r_2'=0$ is fulfilled. For $r_2(\tau) \neq constant$, this implies $\tau=v_2$. The latter relation, once $v_2$ is expressed as $v_2=v_2(r_{10}, r_{2}, r_{30}, r_{40})$ according to Eq. (\ref{v2}), becomes a nonlinear equation in the only unknown $r_2(\tau)$, which can be solved to find the RW.
The velocity $\tau_1 \equiv \tau_{linear}$ and $\tau_2 \equiv \tau_{soliton}$ of the leading and trailing edges of the DSW correspond to the edges of this rarefaction wave and can be calculated from the following limits (see Fig. \ref{S7})
\begin{eqnarray}
\tau_1 &=& \lim_{r_2 \rightarrow r_R^-} v_2(r_{10}, r_{2}, r_{30}, r_{40}) = \nonumber \\
&=& \frac{2r^-_R + r^+ + r^-_L}{4}+\frac{(r^+ -r^-_R)(r^-_L-r^-_R)}{2r^-_R-r^+ - r^-_L};  \label{DSW1}\\
\tau_2 &=& \lim_{r_2 \rightarrow r_L^-} v_2(r_{10}, r_{2}, r_{30}, r_{40}) = \frac{r^-_R + r^+ + 2r^-_L}{4}. \label{DSW2}
\end{eqnarray}
To obtain final formulas that depend only on the quiescent states $r^{\pm}_{L,R}$, we observe that, across the shock, one has a constant $r^+=r^+_L$, as clear from Fig. \ref{S6}. 

Moreover, since the minimum power of the cn-oidal wave turns out to be $\rho_{min}=(r_1-r_2-r_3+r_4)^2/16$ \cite{EL1995PhysD,Hoefer2006PRA}, the existence in the DSW of a cavitating or vacuum state $\rho_{min}=0$ requires $r_2=r_1-r_3+r_4$. Therefore, in analogy with Eqs. (\ref{DSW1}-\ref{DSW2}), we obtain the self-similar location of the vacuum by performing the following limit
\begin{eqnarray} \label{vacuum}
\tau_0 &=& \lim_{r_2 \rightarrow r^-_R - r^-_L + r^+_L} v_2(r_{10}, r_{2}, r_{30}, r_{40}) =\\
&=& \frac{r^+_L + r^-_R}{2} - \frac{r^-_L - r^+_L}{2} \left[1 -  \frac{r^-_R - r^-_L}{r^+_L  + r^-_R - 2 r^-_L} \frac{E(m)}{K(m)} \right]^{-1} \nonumber
\end{eqnarray}
where $m = \left(  \frac{r^-_L - r^+_L }{r^-_R - r^-_L }\right)^2$.
Finally, we obtain Eqs. (5) and (6) in the letter by substituting in Eqs. (\ref{DSW1}-\ref{DSW2}) and (\ref{vacuum}), respectively, the boundary value of Riemann invariants expressed in terms of $\rho$,
i.e. $r^{\pm}_{L,R}=\pm 2\sqrt{\rho_{L,R} }$.


\begin{thebibliography}{99}
\bibitem{rho} Clearly, choosing $P_R$ as the reference power amounts to fix $\rho_R=1$. However, in order to write more general formulas we leave both $\rho_R$ and $\rho_L$ as free boundary values.
\bibitem{warnvelocity} $\tau=t/z$ is referred to as a velocity, as commonly used in shock wave theory, though, dimensionally, this is an inverse velocity.
\end{thebibliography}
\end{document}